# CZT imaging detectors for *ProtoEXIST*

J. Hong[a]*, J. E. Grindlay[a], N. Chammas[a], A. Copete[a], R. G. Baker[b], S. D. Barthelmy[b], N. Gehrels[b], W. R. Cook[c], J. A. Burnham[c], F. A. Harrison[c], J. Collins[d] and W. W. Craig[d]

[a]Harvard Smithsonian Center for Astrophysics, 60 Garden St., Cambridge, MA 02138,
[b]NASA Goddard Space Flight Center, Greenbelt, MD 20771
[c]California Institute of Technology, Pasadena, CA 91125
[d]Lawrence Livermore National Laboratory, Livermore, CA 94550

## ABSTRACT

We describe the detector development for a balloon-borne wide-field hard X-ray (20 – 600 keV) telescope, *ProtoEXIST*. *ProtoEXIST* is a pathfinder for both technology and science of the proposed implementation of the Black Hole Finder Probe, Energetic X-ray Imaging Survey telescope (EXIST). The principal technology challenge is the development of large area, close-tiled modules of imaging CZT detectors (1000 cm² for *ProtoEXIST1*). We review the updates of the detector design and package concept for *ProtoEXIST1* and report the current development status of the CZT detectors, using calibration results of our basic detector unit – 2 x 2 x 0.5 cm CZT crystals with 2.5 mm pixels (8 x 8 array). The current prototype (Rev1) of our detector crystal unit (DCU) shows ~4.5 keV electronics noise (FWHM), and the radiation measurements show the energy resolution (FWHM) of the units is 4.7 keV (7.9%) at 59.5 keV, 5.6 keV (4.6%) at 122 keV, and 7.6 keV (2.1%) at 356 keV. The new (Rev2) DCU with revised design is expected to improve the resolution by ~30%.

**Keywords:** CZT detector, hard X-ray telescope, spectral response

## 1. INTRODUCTION

*ProtoEXIST* is a balloon-borne wide-field hard X-ray (20 – 600 keV) telescope and a pathfinder for both science and technology of the Energetic X-ray Imaging Survey telescope (EXIST) [1,2]. As the proposed implementation of the Black Hole Finder Probe, EXIST employs a coded-aperture imaging technique with CZT detectors to capture the increasingly interesting hard X-ray sky. In order to meet its ambitious scientific goals, EXIST requires a very large array of detector area (~6 m²) with relatively fine detector pixel size (~1.2 mm), good energy resolution (<3 keV) and high sensitivity over the ~10 – 600 keV band. Therefore the principal technology challenge is the development of large area, close-tiled modules of imaging CZT detectors that can operate under the limited resources available in the space mission (e.g. <100 μW/pixel power consumption). We plan to demonstrate the feasibility of such technology through *ProtoEXIST* by conducting balloon flight tests under near space conditions. In *ProtoEXIST1*, we will build 1000 cm² CZT detector arrays with 2.5 mm pixel and in *ProtoEXIST2*, we will refine the pixel size to 1.2 mm. In this paper, we review our detector design and package concept and report the current development status of the program using our basic detector units.

## 2. DETECTOR CONCEPT FOR *PROTOEXIST1*

Hong et al. (2005) outlined the initial detector plane design concept for *ProtoEXIST1* [3]. Since then, we have made a few changes to improve our packaging concept to resolve practical concerns that became apparent during development. Here we review the current concept of the detector plane design.

The basic building block of the detector plane is a Detector Crystal Unit (DCU). A DCU is made of a 2 x 2 cm CZT crystal bonded on an interposer board (IPB), which interconnects the 2-D 8x8 arrays of anode pixel pads on the bottom of the CZT crystal to a 1-D array of 64 input pads for an ASIC (for signal handling) mounted on the backside of the IPB (Fig. 1). In order to cover the wide energy range (10 – 600 keV), we use ≥5mm thick CZT crystals. The RadNet ASICs, derived for a Homeland Security application based on ASIC originally developed for HEFT [4,5], are selected suitable for *ProtoEXIST1* because of its approximate gain (~20 – 1000 keV), right form factor for 2.5 mm-pixel detectors, multi-pixel pulse-profile readout capability (see section 4) and low power consumption (<100 μW/pixel). A few dozen DCUs have been built over the past years (Rev1), and they have been tested on a prototype test board, the "poptart" board. Their performance is well understood (see section 3) and a new revision (Rev2) has been made to the IPB design accordingly, and we are expecting the next batch (Rev2) of IPBs and assembled DCUs in September, 2006.

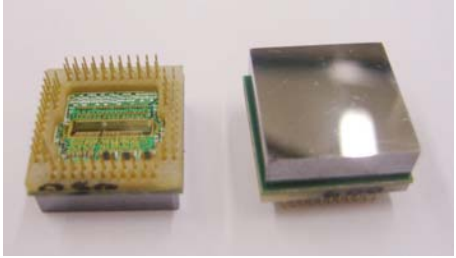

Fig. 1 Detector Crystal Units (DCUs): 2 x 2 x 0.5 cm CZT Crystal + Interposer Board + RadNet ASIC.

A 2 x 4 array of DCUs are close-packed onto a Detector Crystal Array (DCA). A DCA thus consists of two stacked circuit boards, covering 4 x 8 cm detector area (Fig. 2). The top board of a DCA contains matching 8 female sockets for DCA mounting and the bottom board of the DCA contains an Altera* Cyclone II FPGA to process signals from the 8 ASICs on the 8 DCUs. The departure from the original concept for DCA with 2 x 2 DCUs is made to accommodate a resourceful FPGA for signal processing with multiple modes and to allow tight packaging (< 0.5 – 1 mm gap between crystals) as well as relatively easy mounting/dismounting of each module for testing. The FPGA on the DCA will have four independent channels, and each channel handles the data stream from two ASICs. The Verilog code run on the FPGA will allow multiple operation modes, taking a full advantage of the ASIC features (see section 4). The first batch of the DCA boards has just been built, and we plan to test these boards through a simple test board shortly.

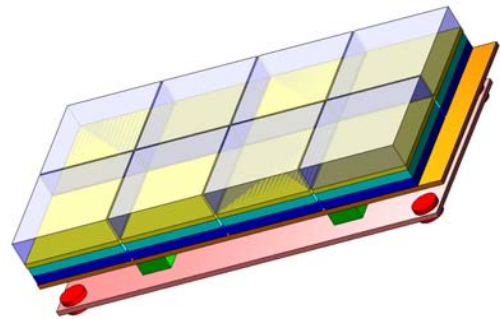

Fig. 2 Detector Crystal Array (DCA, 32cm²): 2x4 DCUs + top socket board for DCUs + bottom FPGA board

The final stage of building a detector plane for *ProtoEXIST1* is the Detector Module (DM). A DM consists of 2 x 4 arrays of DCAs mounted on a mother board, the FPGA controller board (FCB), covering 256 cm² of X-ray collecting area (Fig. 3). The FCB hosts another FPGA of the Altera Cyclone II family to control and process signals from 8 DCAs, and the FCB communicates with the flight computer by Ethernet protocol. The FPGA on the FCB will have 32 independent channels matching 4 x 8 channels of 8 FPGAs in 8 DCAs. The FCB also has two sockets for two high voltage power supply (HVPS) boards that are vertically mounted (yellow boards in Fig. 3) on each side of the DM. Currently we are designing the FCB according to the DCA board specification.

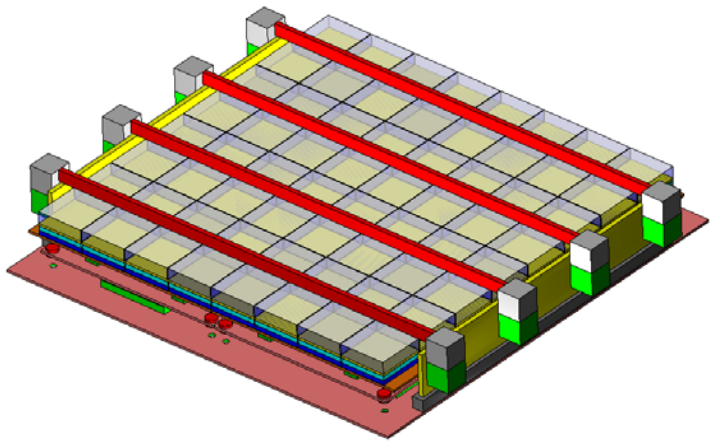

Fig.3 Detector Module (DM, 256 cm²): 2x4 DCAs + Motherboard with an FPGA + 2 HV PS Boards

When an event occurs, the ASIC discriminator output alerts the corresponding channel of the FPGAs in both the DCA and the FCB. Then, the DCA FPGA initiates the data readout sequence on the ASIC, while the FCB FPGA grabs the proper time tag and readout the shield status. When the ASIC readout is finished, the FCB FPGA will combine the X-ray data packet from the DCA FPGA with the accessory data packet, and store them in the buffer that will be transferred via Ethernet to the flight computer for on-board processing and storage and transmission to the ground.

---

*ProtoEXIST1* consists of 4 independent telescopes, and each employs an identical DM, for 1024 cm² detector area in total. In order to characterize the detector performance in a near space environment, we plan to configure the 4 telescopes differently providing two fields of view options (fully coded: 9° vs. 18°), two shielding configurations (plastic+passive vs. active+passive), and two different mask types (random vs. uniformly redundant array).

## 3. DCU PERFORMANCE MODELLING

Fig. 4 shows the test setup for DCUs. It is a single board system ("poptart" board) that can control 4 DCUs through an Altera Cyclone I FPGA. It can communicate with a PC through a USB or serial port. The board contains a low-noise, low-power HVPS circuit for biasing crystals. Since the FPGA on DCAs will use the same Altera cyclone family, the poptart board has been also used to develop the necessary FPGA code for several different operation modes.

The current batch (Rev1) of DCUs contains CZT crystals from eV products[*] and Orbotech Ltd[†] (formerly known as IMARAD). These crystals are bonded on an IPB using a low temperature solder bonding technique developed at Aguila Technologies[‡]. The IPB is made of the FR4 at DDi[§], and the components are mounted at Aguila Technologies.

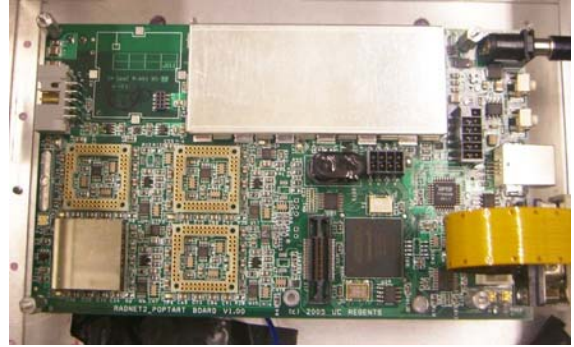

The major noise components or contributors are the ASIC intrinsic noise, the input capacitance due to the wire bond, the traces in the IPB, and the bonding material, and the leakage currents when a biasing HV is applied across the crystals. In order to identify these various components of noise sources, we have made a few different kinds of DCUs and we have modeled the noise contribution from each component separately. We use a bare IPB (IPB 041) and DCU (IPB 050) with an Orbotech crystal (Pt/Pt contact, no guard band) as examples in the following. These are a typical unit of their kind, and units of the same kind usually show a similar behavior of electronics noise.

Fig. 4 DCU test setup using a "poptart" board. This can accommodate 4 DCUs and control them through an Altera Cyclone FPGA. The board contains HV PS circuit shielded in an Al case.

First, the ASIC intrinsic noise is well estimated [4,5] and it can be directly compared with the pulser test results on bare IPBs (no crystal) after severing the connection of the ASIC input wire bonds. Second, there is the noise contribution due to the input capacitances of the IPB. Fig. 5a shows the input trace lengths of the IPB in pixel coordinates, which indicates the top two rows of pixels have longer traces than other pixels. This variation is due to packaging constraints in the IPB. Using the simplified trace geometry of the IPB, we estimated the associated parasitic capacitance by FASTCAP[**] (Fig. 5b). As expected, the pattern of capacitance estimates and the trace lengths matches fairly well. Due to the simplification of the geometry model (e.g. no cross talk between traces), the absolute value of the capacitance estimates can be somewhat off by a constant factor (<2x), but the pattern among various pixels is likely valid.

Fig. 5c shows the FWHM of the electronics noise from pulser tests on a bare IPB (IPB 041). The noise pattern among pixels matches with our capacitance estimates, which indicates the pixel dependent noise variation of the bare IPB is mainly due to the variation of trace lengths of the IPB. Once a crystal is mounted on the bare IPB, there will be some additional input capacitance due to the bonding material. Fig. 5d shows the similar pulser test results on a DCU with a crystal (IPB 050), indicating roughly 1.3 keV increase in the noise compared to the bare IPB. We estimate the bonding material and the underfill are responsible for this noise increase and together provide ~1pF additional capacitance to the total input capacitance of the ASIC.

Third, the leakage currents introduced when a HV bias (– 600 V) is applied across the crystal, will contribute to the additional electronics noise. This noise will be added quadratically to the other noise since they are more or less independent. Fig. 5e shows the electronics noise (FWHM) of the same DCU module (IPB 050) at – 600V. The overall

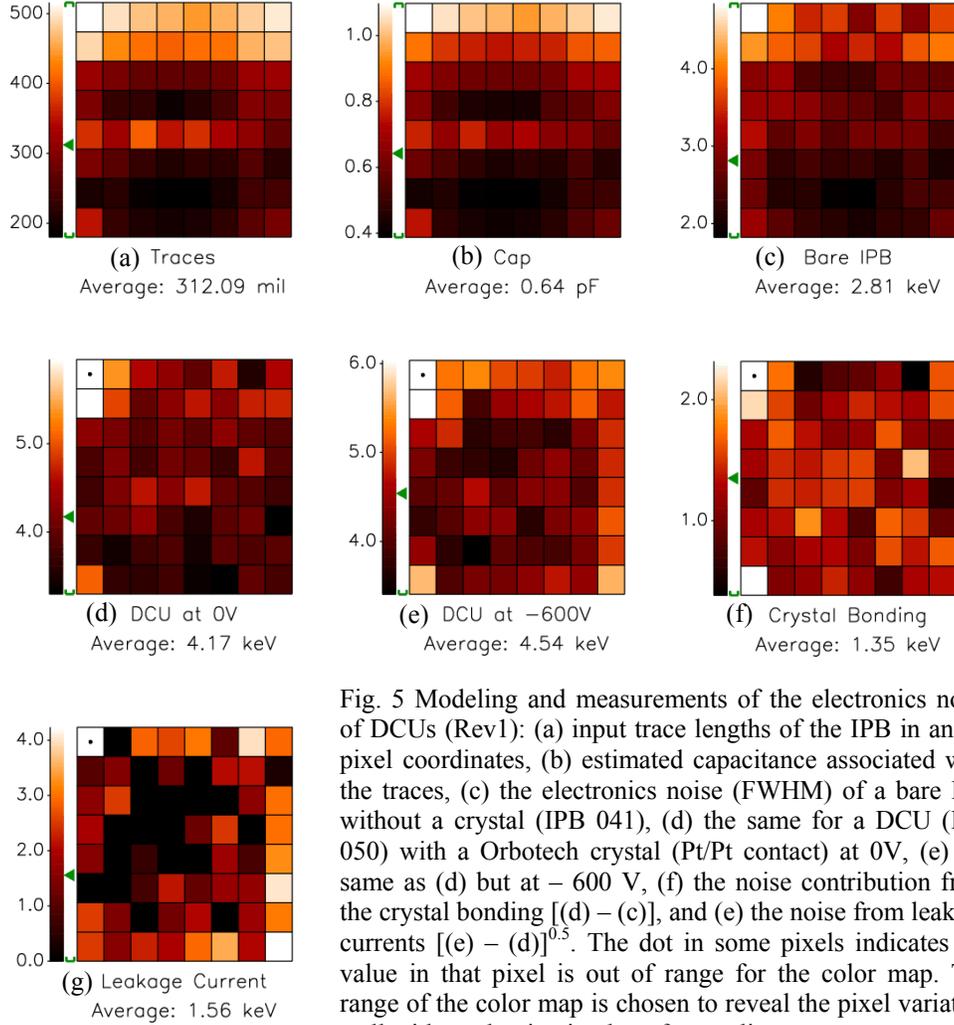

**(a)** Traces
Average: 312.09 mil

**(b)** Cap
Average: 0.64 pF

**(c)** Bare IPB
Average: 2.81 keV

**(d)** DCU at 0V
Average: 4.17 keV

**(e)** DCU at −600V
Average: 4.54 keV

**(f)** Crystal Bonding
Average: 1.35 keV

**(g)** Leakage Current
Average: 1.56 keV

Fig. 5 Modeling and measurements of the electronics noise of DCUs (Rev1): (a) input trace lengths of the IPB in anode pixel coordinates, (b) estimated capacitance associated with the traces, (c) the electronics noise (FWHM) of a bare IPB without a crystal (IPB 041), (d) the same for a DCU (IPB 050) with a Orbotech crystal (Pt/Pt contact) at 0V, (e) the same as (d) but at − 600 V, (f) the noise contribution from the crystal bonding [(d) − (c)], and (e) the noise from leakage currents $[(e) - (d)]^{0.5}$. The dot in some pixels indicates the value in that pixel is out of range for the color map. The range of the color map is chosen to reveal the pixel variation well without domination by a few outliers.

noise pattern has also changed under the HV bias, and the pattern indicates the excess of leakage currents (and thus noise) at the edge pixels. This is expected because of lack of any guard band around the crystal, which is usually applied for Orbotech crystals with blocking contacts to reduce relatively higher leakage currents at the edge pixels. The overall noise increase from the case with no bias is about 0.3 keV, which means the leakage currents alone generate about $1.0 - 1.5$ keV FWHM noise. This corresponds about $1.0 - 1.5$ nA/pixel leakage currents, which matches with properties of other Orbotech crystals with blocking contacts and no guard band [6].

Table 1 summaries the electronics noise of the current batch (Rev1) of DCUs. The listed four major components contribute more or less similar ($\sim 1.0 - 1.5$ keV) to the total 4.5 keV FWHM noise, and their estimate and experimental tests using pulser data match very well.  Based on this analysis, we have revised the design of IPBs and the assembly method of DCUs to reduce the noise (Rev2). In the case of IPBs, we changed the board material from FR4 to Arlon 55NT for a smaller dielectric constant ($4.5 \rightarrow 3.8$), and we also reduced the trace width from 5 to 3 mil. The FASTCAP simulation predicts about 30% reduction in the input capacitance of IPBs from these changes. For crystal bonding, we plan to minimize the underfill, effectively reducing dielectric constant around the bonding metal from 4.5 to 1.0, which will reduce the capacitance of the bonding material by a factor of 4. As for leakage currents, we plan to use crystals from

Redlen Technologies[*] for the next batch (Rev2). Our preliminary I-V measurements of a sample of Redlen crystals show <0.5 nA/pixel at − 600V. As the result of these changes, we expect the electronics noise (FWHM) of the Rev2 DCUs will be ~3.0 keV. The first 50 Rev2 DCUs are currently being fabricated.

As for *ProtoEXIST2* and EXIST, we will either remove the IPB by direct flip-chip bonding between crystals and ASICs with 2-D matching input pad arrays, or at worst using 2-D to 2-D de-magnifying IPBs of minimal traces, so that the input capacitance of the IPB will be negligible and there will be no contribution from the wire bonds of ASICs. In the case of using the same kind of crystals, the leakage current per pixel will be about ¼ of that in the Rev2 DCUs because of the smaller pixel size (¼ in area). Therefore, we expect ≤2.0 keV electronics noise from the DCUs for *ProtoEXIST2* and EXIST.

Table 1 Simplified Noise Breakdown (FWHM in keV)

|  | Estimates for Rev1 DCUs | Measurements (IPB 041, 050) | Estimates for Rev2 DCUs | DCUs for *ProtoEXIST2* & EXIST |
|---|---|---|---|---|
| AISC intrinsic + wire bond | 1.5 − 2.0 | 1.5 − 2.0 | 1.5 − 2.0 | 1.5 − 1.7 (flip chip bond) |
| Input traces | 2.5 − 3.5 (1.0 − 1.5) | 2.8 | 2.2 − 3.0 (0.7 − 1.0) | 1.5 − 1.9 (0.0 − 0.2) |
| Crystal bonding | 3.5 − 5.0 (1.0 − 1.5) | 4.2 | 2.5 − 3.5 (0.3 − 0.5) | 1.8 − 2.4 (0.3 − 0.5) |
| Leakage current (HV: −600V) | 3.6 − 5.0 (1.0 − 1.5) | 4.5 | 2.5 − 3.6 (0.5 − 0.7) | 1.8 − 2.4 (0.1 − 0.2) |

The numbers indicate the accumulated noise and the values in ( ) indicate the contribution from each component.

## 4. MULTI-PIXEL PULSE PROFILE READOUT

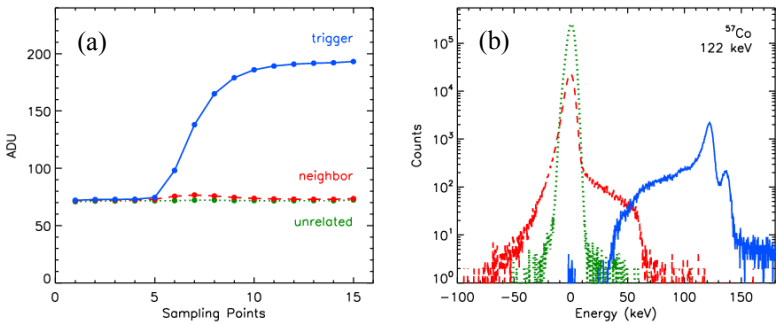

Some of the key advantages of using RadNeT ASICs for *ProtoEXIST1* are first, multi-pixel readout and second, the pulse profile sampling. For every event, the ASIC can record signals from any combination of pixels, which can be useful for charge split events arising from charge spreading or Compton scatterings. In addition to multi-pixel readout, the ASIC provides 16 sampling points of the signal profile for each pixel [4]. Fig. 6 shows the averaged pulse profile from triggered pixels, 8 neighbors, and the rest of the pixels. The rise time information given by pulse profile can be used for depth sensing and correcting for incomplete charge collection. To measure pulse height for a pulse profile, we average the first and last 4 or 5 samples, and take the difference, which turns out to work well. At maximum (calibration mode, see below), one can record 16 samples x 64 pixels =1024 numbers for a given event.

Fig. 6 Averaged pulse profiles (a) and the pulse height histograms (b) from a $^{57}$Co source: the (blue) solid lines for triggered signals, the (red) dash lines for neighbor signals and the (green) dotted lines for signals from the rest of pixels. Note that the 16th sampling points are not used.

The downside of this full data recording mode is the deadtime constraint on the count rate, which is limited to about 50 counts/sec/DCU. While this limit exceeds the requirements for *ProtoEXIST* and EXIST (the nominal count rate <~0.5 counts/cm²/sec = 2.0 counts/sec/DCU), the full data recording is usually unnecessary for cosmic X-ray events. Since the count rate limit is inversely proportional to the number of pixels to read out per event, we need to identify the most efficient and useful data readout modes. Tentatively we plan to implement multiple operation modes in the FPGA: debugging mode (read out one designated pixel per event), variable mode (read out all triggered pixels), normal (variable mode + the four nearest neighbor pixels), full (variable mode + eight nearest neighbor pixels), and calibration mode (readout all 64 pixels). In order to identify the most efficient and useful data taking mode, we have performed a series of radiation measurements in the calibration mode. In the following, we review the spectral analysis procedure using the calibration data taken from an $^{241}$Am, $^{57}$Co, and $^{133}$Ba source.

Fig. 6b shows the pulse height histogram from the $^{57}$Co source. The (blue) solid line is the histogram from the triggered pixel, the (red) dotted line from the 8 nearest neighbors and the (green) dotted histogram from the rest of pixels. The (blue) solid line reveals peaks from 122 and 136 keV X-ray lines from the source. The (green) dotted histogram indicates the electronics noise distribution since the non-collecting pixels far from the trigger pixel measures nothing but the random electronics noise. The noise distribution is well confined within −10 to 10 keV, which is consistent with the FWHM of the electronics noise being 4.5 keV. Note that the actual threshold of the system has to be set at around 30 keV or higher in order to avoid the large number of noise triggers. This is because large base line fluctuations would allow the <10 keV noise to possibly appear above the threshold to the ASIC discriminator that cannot perform the averaged based line subtraction possible with off-line analysis.

The spectra (Fig. 6b) from neighbor pixels show two interesting features different from the noise spectra. The shoulder on the positive side (>10 − 15 keV) is due to charge splitting events arising from either charge spreading or Compton scattering. The shoulder on the negative side (< − 15 keV) indicates the negative induced charges in the neighboring pixels due to the depth-dependent hole trapping [7]. One can use this negative signal from neighbors to improve the spectral response of the detector. We have developed techniques to allow for these effects.

First, to correct for charge splitting events, we use the strongest signals among 4 nearest neighbors vs. the signals from the triggered pixels in Fig. 7a. One can see the spots from 122 and 136 keV lines in the plot, and the diagonal streak extended from the 122 keV (and 136 keV) spot indicates the charge splitting events. Since the noise drops at around 10 keV, we simply add the signal greater than 15 keV to the signal from the triggered pixel for the correction.

Second, to correct for incomplete charge collection, we use the negative signals induced in neighboring pixels due to the trapped holes [7]. We use the phase space of the signal sum from 4 nearest neighbors vs. the signals from the triggered pixel (Fig. 7b) since the negative pulse is induced in all nearest neighbor as opposed to the charge split events that occur mostly between the triggered pixel and one neighbor pixel. We identify the strong correlation of the negative side of the neighbor pulse sum with the low energy tail of the triggered signals from the 122 keV line, using $2^{nd}$ order polynomial function (Fig. 7c). Based on the identified correlation, one can add the missing charge for every event in the low energy tail of the 122 keV peak. This correction has to be done pixel by pixel (the plots in Fig. 7 are from all 64 pixels) and energy by energy to be most effective. In the analysis here, the correction is done independently for three lines: 60, 122 and 356 keV. Fig. 7d shows the same phase space after both corrections for charge split and incomplete charge collection. Fig. 7e and 7f show the events from a $^{133}$Ba source before and after the corrections, which reveals more dramatic changes at high energies.

Fig. 8 and Table 2 summarize the spectral response before and after the corrections using combined data of all 64 pixels. In Fig. 8, the (black) dashed lines are the raw histograms, the (blue) dotted lines after the correction only for the charge splitting events, and the (red) solid lines after both corrections for charge split and incomplete charge collection. Fig. 8 and Table 2 contain three numbers for each spectrum to measure its spectral response. First, the FWHM (in keV) is calculated from a simple Gaussian fit on the upper half of the histogram around the peak. This FWHM misses the low energy tail of the peak, so another FWHM is estimated from the Gaussian plus exponential fit covering the whole peak. Finally, the photo-peak efficiency is calculated from the fractional count ratio between the peak region and threshold to peak. The photo-peak efficiency here is used as a loose indicator for the spectral quality, but not intended for the precise measurement (see Table 2 for the range). For example, in the case of the 356 keV line from the $^{133}$Ba source, the current definition of the photo-peak efficiency underestimates the true efficiency since the current definition does not separate contributions from other X-ray lines.

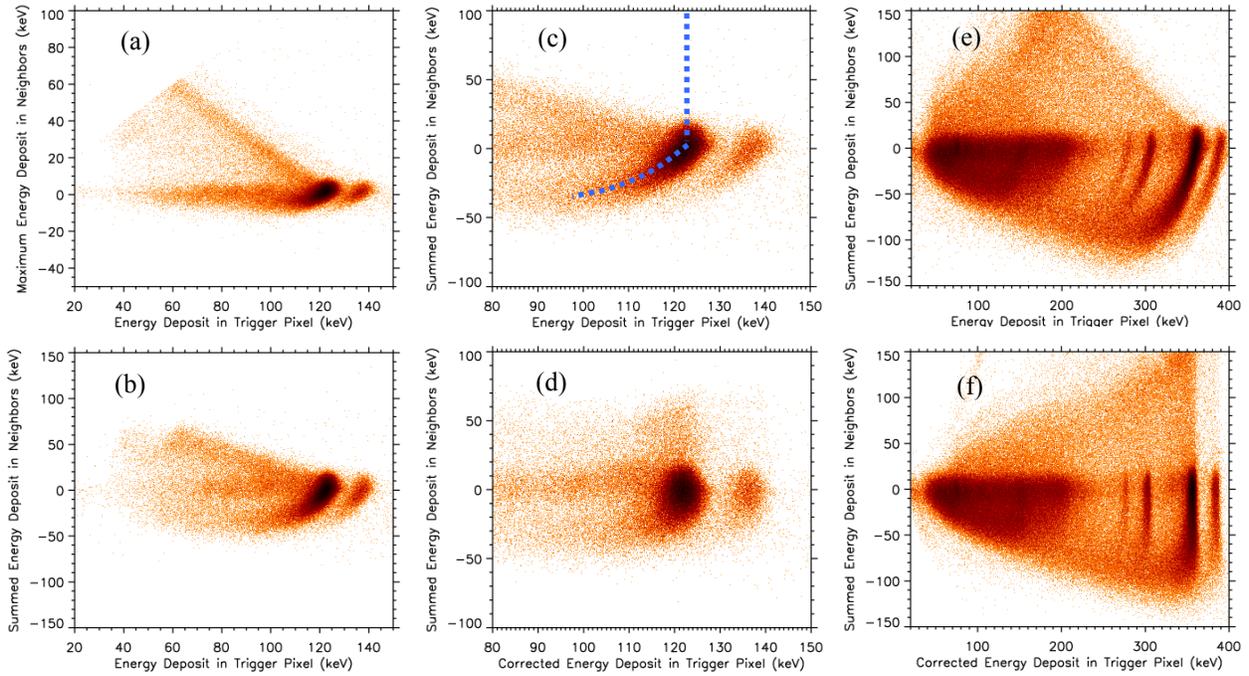

Fig. 7 Correcting charge split and incomplete charge collection: (a) the largest signals among four nearest neighbors vs. the triggered signals before any correction from a $^{57}$Co source, (b) the summed signals of four nearest neighbors vs. the triggered signals before any correction , (c) the same as (b) but zoomed in at around 122 keV along with the correlation track, (d) the same as (c) after the corrections (see the text) pixel by pixel, (e) the same as (c) but from a $^{133}$Ba source, and (f) the same as (e) after the correction pixel by pixel.

The correction for charge splitting events improves the photo-peak efficiency, but it is not so useful to improve spectral resolution. However, the correction for the incomplete charge collection improves both the resolution and photo-peak efficiency, and the quality of corrections improves with the energy since high energy X-rays have a higher chance to interact deep in the detector. Since the incomplete charge collection is proportional to the amount of trapped holes and thus the interaction depth, one can get the depth information from the relative strength of negativity in the neighbors' signals as well. In case of 60 or 122 keV X-rays, the interactions mostly occur near the cathode surface for 5 mm thick crystals, so that the depth sensing from the rise time of the triggered pulses is not useful due to their >1 mm depth resolution near the cathode surface. Therefore, the correction for incomplete charge collection based on the negative neighbor pulse outperforms the correction based on the rise time of the triggered pulses.

After both corrections, the low energy tail from each spectrum is greatly reduced, and the FWHM resolution is 4.71 (7.92%), 5.58 (4.57%) and 7.60 (2.13%) keV for 59.5, 122 and 356 keV lines. If this trend holds, we expect ~ 1% FWHM at 662 keV, which is quite remarkable for CZT. In the case of the $^{133}$Ba data, note that the relative weakness of the 80 keV line is due to a lead shield set in front of the source to suppress the low energy lines. This is done to efficiently capture the high energy lines under the count rate limitation of the calibration mode.

Fig. 9 shows the 2-D map of the FWHM by a simple Gaussian fit on the corrected histogram from each pixel and the photo-peak efficiency. Note that the average value of the FWHMs (e.g. 4.72 keV for $^{241}$Am) is slightly different from the FWHM of the summed spectra (e.g. 4.71 keV for $^{241}$Am) due to the non-linear nature of the spectral fitting procedure and the pixel variation of the total counts per pixel. Both energy resolution and the photo-peak efficiency show somewhat large pixel to pixel variations (from ±10% to ±30%). In the case of the energy resolution, the variation is due to the pixel variation of the electronics noise described in section 3 and the non-uniformity of the crystal (e.g. leakage current). As we reduce the overall electronics noise the next batch of DCUs (Rev2), and we use crystals of more uniform properties, we expect the pixel variation of the energy resolution will decrease.

The pixel variation of photo-peak efficiency is similar to that of typical Orbotech crystals (with blocking contacts) we have tested in the past. Interestingly the photo-peak efficiency map shows the edge pixels are progressively less efficient with energy. This is the result of missing neighbors (more important at high energies), which reduces a chance of the correction for the incomplete charge collection. Although the charge split events due to charge spreading will not be shared among neighboring crystals, the trapped holes in a crystal may induce detectable negative signals in the neighboring crystals since our detector design has only <1mm gap between crystals. In order to improve the relatively low photo-peak efficiency at edge pixels, we might implement a data taking mode to record signals from neighboring pixels even across neighboring detectors (DCUs or DCAs).

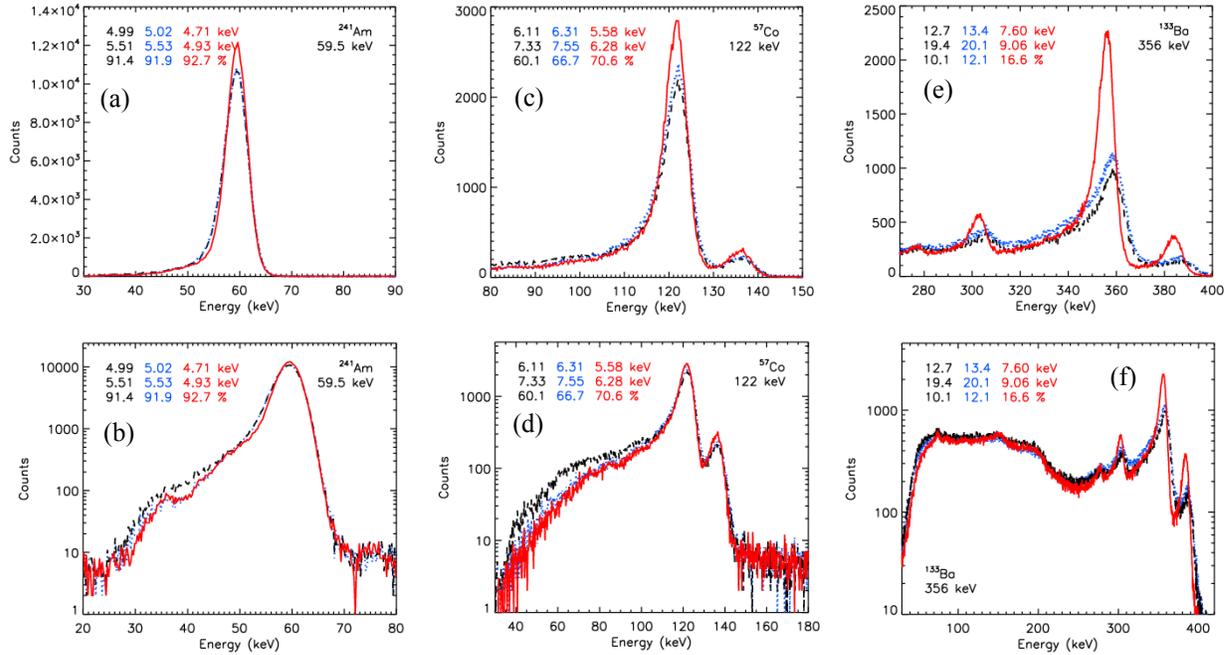

Fig. 8 Comparison of the pulse height histograms before and after the corrections: (red) solid after both corrections, (blue) dotted after only charge split correction, (black) dash for the raw. (a,b) is for $^{241}$Am, (c,d) for $^{57}$Co, and (e,f) for a $^{133}$Ba source. See Table 2.

Table 2 Spectral resolution (FWHM in keV) and photo-peak efficiency (%)

| | 59.5 keV from $^{241}$Am | | | 122 keV from $^{57}$Co | | | 356 keV from $^{133}$Ba | | |
|---|---|---|---|---|---|---|---|---|---|
| | Raw | Charge Split Corr. | Both Corr. | Raw | Charge Split Corr. | Both Corr. | Raw | Charge Split Corr. | Both Corr. |
| FWHM by a Gaussian fit | 4.99 | 5.02 (+0.6%) | 4.71 (−5.6%) | 6.11 | 6.31 (+3.3%) | 5.58 (−8.7%) | 12.7 | 13.4 (+5.5%) | 7.60 (−40%) |
| FWHM by a Gaussian+exp tail fit | 5.51 | 5.53 (+0.3%) | 4.93 (−11%) | 7.33 | 7.55 (+3.0%) | 5.28 (−28%) | 19.4 | 20.1 (+3.6%) | 9.06 (−53%) |
| Photo-peak efficiency [Energy ranges for PPE] | 91.4 | 91.9 (+0.5%) | 92.7 (+1.4%) | 60.1 | 66.7 (+11%) | 70.6 (+17%) | 10.1 | 12.1 (+20%) | 16.6 (+64%) |
| | 53 − 65 keV / 40 − 70 keV | | | 114 − 129 keV / 40 − 129 keV | | | 345 − 365 keV /40 − 370 keV | | |

The numbers in ( ) indicate the changes in %, compared to the raw histograms. The Gaussian + exp tail function is given by $A_0 \exp(-\Delta E^2/A_2) + A_3 \exp(1 - \Delta E/A_4) [1 - \exp(-\Delta E^2/A_5)]$ where $\Delta E = E - A_1$ and $A_3 = 0$ when $\Delta E > 0$. Both Corr. is after both corrections for charge split and incomplete charge collection.

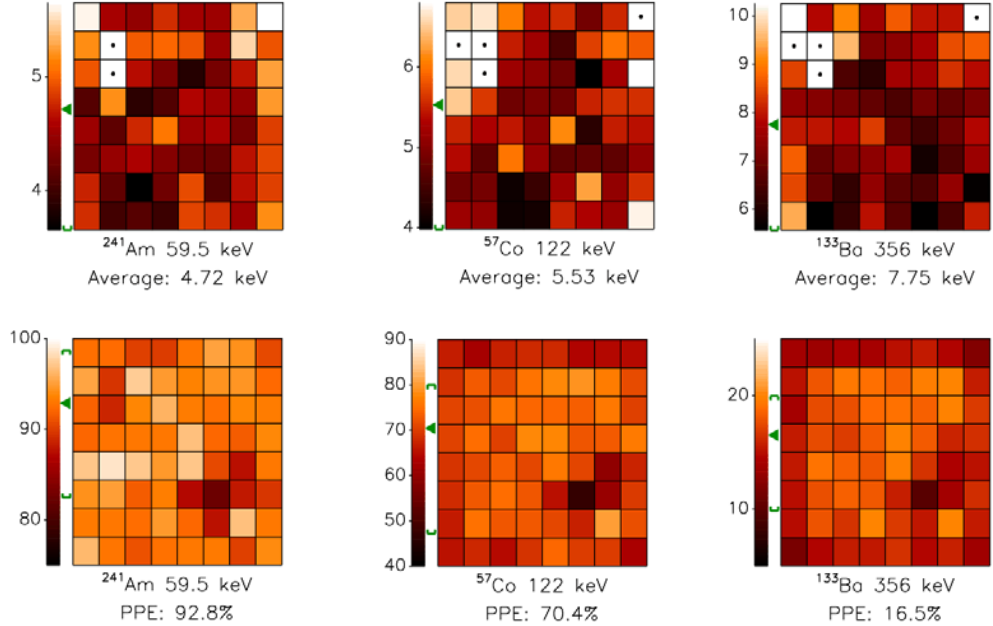

Fig. 9 The energy resolution (FWHM, top) and the photo-peak efficiency (PPE, bottom) from [241]Am (left), [57]Co (middle) and [133]Ba (right).

## 5. ADDITIONAL PROCESSING AND DEPTH SENSING

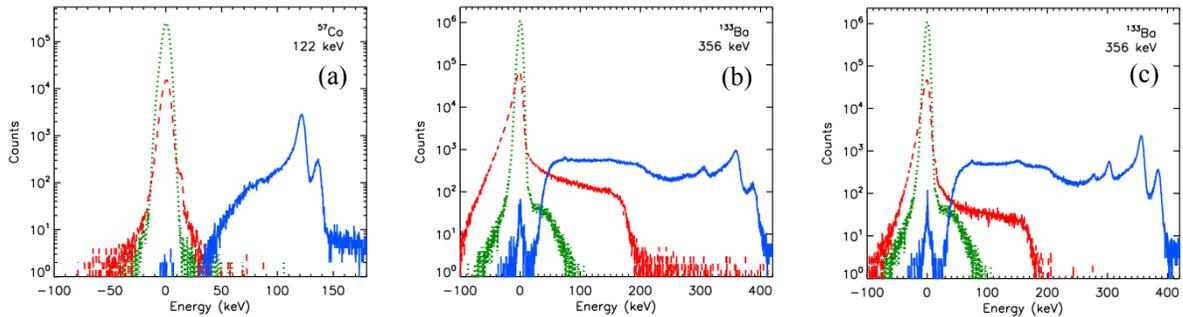

Fig. 10  Spectral decomposition of pulse height histograms from the [57]Co source after correction (a; see Fig. 6b for before correction), the same from the [133]Ba source before correction (b) and the same after correction (c): the (blue) solid lines for triggered signals, the (red) dashed lines for neighbor signals and the (green) dotted lines for signals from the rest

We have kept track of each case in the pulse height histograms of neighbors' signals during the corrections so that we can tag it out if it is used for correction. Fig. 10 shows the spectral composition of the [57]Co and [133]Ba data before and after correction. In the case of the [57]Co data, the correction makes the neighbor histogram resembles the noise histogram, which indicates the correction is almost complete. But there is still room for improvement. For example, in the noise histogram (green) of Fig. 10a (and Fig. 6b), the small excess in the left wing ($< -10$ keV) compared to the right wing ($>10$ keV) indicates some of the non-collecting pixels (likely the next nearest neighbors) can also record negative induced charges.

In the case of the $^{133}$Ba data, the large difference between the neighbor and noise histograms after the corrections indicates there are still much more to correct. In addition, the noise histogram indicates the charge split events or the negative induced signals seem to extend the next nearest neighbors. This is possible when an event undergoes both Compton scattering and charge spreading from the subsequent interaction. Finally the true multi-hits from Compton scattering events are not corrected in the above spectra.

The charge split between the triggered pixel and the nearest neighbors can come from either Compton scattering or charge spreading. In order to see only Compton scattering events, we have selected the events with two interactions (triggered pixels) that occur at least two pixels apart (note: this will be more important for the smaller pixels of *ProtoEXIST2* and *EXIST*). Fig. 11a shows the (red and blue) histograms of each interaction and the histogram of the summed signal. The histograms of each interaction do not resemble the spectra from the $^{133}$Ba source, but the histogram of the summed signal reveals the correct $^{133}$Ba spectrum, which indicates the successful reconstruction of the Compton scattering events. Fig. 11b shows the histogram of events with single or multiple triggers within one pixel apart (blue), and the histograms of summed signals from events with two (red) and three (green) interactions that occur at least two pixels apart.

For a given pixel, the weighting potential extrudes the neighbors. This results in a transient signal in neighbors when the electron clouds move down to the anode pad of the collecting (triggering) pixel. The strength of the transient signal is proportional to the size of the electron clouds and the proximity to the neighbors, which allows a sub-pixel identification of the interaction point [8]. In our pulse profiles, we do see these transient signals in the neighbors, and even the average profile in Fig. 6a shows a small excess of the neighbor pulse profile with respect to the rest of the pixel at the rising time of the triggered pixels. This information allows another set of the analysis and refinement techniques.

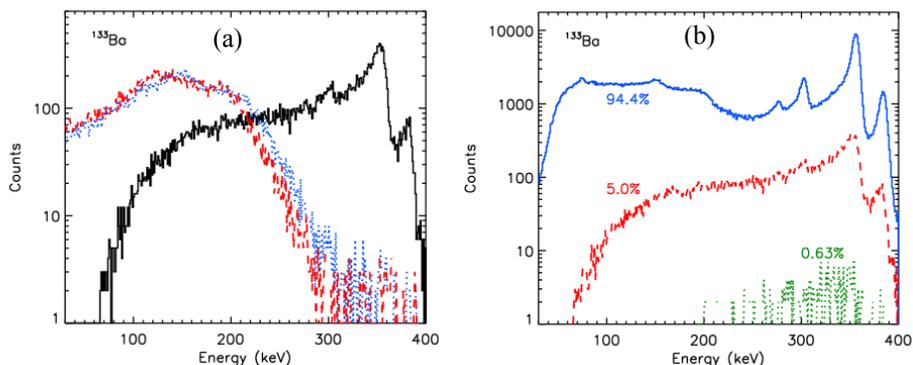

Fig. 11 (a) Reconstructing Compton scattering events from the selected ($^{133}$Ba) events with two interactions that occurred at least two pixels apart: histograms of signals from each interaction (red dashed and blue dotted lines) and summed signal (black solid line). (b) Relative contributions of the events with two (red dashed line) and three interactions (green dotted line) at least two pixels apart.

Finally for low energy X-ray interactions ($< \sim 150$ keV) near the cathode side, the rise time of the pulse profile may not be as useful as the negative signal in neighbors, but for high energy X-ray interaction occurring deep in the detector, the rise time can provide the additional meaningful information on the depth and correction for incomplete charge collection. Therefore, one can combine the information from both the negative neighbor signal and the rise time to improve the correction for incomplete charge collection as well as depth measurement. Judging from the strong correlation in the phase diagram of Fig. 7c & 7e and the similar diagram using cathode depth sensing in Hong, et al. (2004) [9], such analysis would also likely provide the $\ll 1$mm depth information over the full range of the depth from the cathode side to anode side for high energy X-rays in 5 mm thick crystals. Further studies using finely collimated radiation [9] are required to verify the depth sensing sensitivity of multi-pixel pulse-profile readout system.

## 5. SUMMARY

We have reviewed the current detector design and package concept for *ProtoEXIST*, which allows a virtually contiguous 256 cm$^2$ area of CZT detector with sub-mm gap between crystals. The electronics noise of the current batch (Rev1) of the DCUs is ~4.5 keV (FWHM), and the noise from the next batch (Rev2) is expected to be reduced by ~30%. From the effective correction for charge split and incomplete charge collection using neighbor signals, we get 4.7, 5.6, and 7.21 keV FWHM resolutions, or 7.9%, 4.6% and 2.1% , for 59.5, 122 and 356 keV lines from the Rev1 DCUs. The efficient correction for spectral response using neighbor signals indicates the normal (reading 4 nearest neighbors in addition to the triggered pixel) or full modes (reading 8 neighbors) are necessary to make use of the full potential of the spectral resolution of these imaging hard X-ray CZT detectors.

## ACKNOWLEDGMENTS

This work is supported in part by NASA grants NAG5-5279, NAG5-5396 and NNG06WC12G. We thank A. Capote at Aguila Technologies for help on IPB packaging.

## REFERENCES

1. J. E. Grindlay et al., "EXIST: mission design concept and technology program," in *X-Ray and Gamma-Ray Telescopes and Instruments for Astronomy*; J. E. Truemper, H. D. Tananbaum; Eds., *Proc. SPIE* **4851**, pp. 331-344, 2003.
2. W. W. Craig, J. Hong and EXIST Team, "The Energetic X-ray Imaging Survey Telescope (EXIST): Instrument Design Concepts," AAS, **205**, 5001C, 2004
3. J. Hong et al., "Detector and telescope development for *ProtoEXIST* and fine beam measurements of spectral response of CZT detectors" in *UV, X-Ray, and Gamma-Ray Space Instrumentation for Astronomy XIV*; O. H. Siegmund; Eds., *Proc. SPIE*, **5898**, pp. 173-181, 2005
4. W. R. Cook, J. A. Burnham and F. A. Harrison, "Low-noise custom VLSI for CdZnTe pixel detectors" in *EUV, X-Ray, and Gamma-Ray Instrumentation for Astronomy IX*; O. H. Siegmund, M. A. Gummin; Eds., *Proc. SPIE* **3445**, pp. 347-354, 1998
5. W. W. Craig, L. Fabris, J. Collins and S. Labov, "A Cellular-Phone Based Radiation Detector: Technical Accomplishments of the RadNet Program," UCRL-TR-215280.
6. S. Vadawale et al., "Multipixel characterization of imaging CZT detectors for hard x-ray imaging and spectroscopy" in *Hard X-Ray and Gamma-Ray Detector Physics VI*; A. Burger, R. B. James, L. A. Franks; Eds., *Proc. SPIE* **5540**, p. 22-32, 2004
7. J. D. Eskin, H. H. Barrett, and H. B. Barber, "Signals induced in semiconductor gamma-ray imaging detectors" J. Appl. Phys. **85**, pp 647-659, 1999
8. T. Narita, J. E. Grindlay, J. Hong and F. C. Niestemski, "Anode readout for pixellated CZT detectors," in *X-Ray and Gamma-Ray Instrumentation for Astronomy XIII*; K. A. Flanagan, O. H. Siegmund; Eds., *Proc. SPIE* **5165**, pp. 542-547, 2004.
9. J. Hong et al., "Cathode depth sensing in CZT detectors" in *X-Ray and Gamma-Ray Instrumentation for Astronomy XIII*; K. A. Flanagan, O. H. Siegmund; Eds., *Proc. SPIE* **5165**, pp. 54-62, 2004